\PassOptionsToPackage{unicode}{hyperref}
\PassOptionsToPackage{hyphens}{url}
\PassOptionsToPackage{dvipsnames,svgnames,x11names}{xcolor}
\documentclass[
]{article}

\usepackage{amsmath,amssymb}
\usepackage{lmodern}
\usepackage{iftex}
\ifPDFTeX
  \usepackage[T1]{fontenc}
  \usepackage[utf8]{inputenc}
  \usepackage{textcomp} 
\else 
  \usepackage{unicode-math}
  \defaultfontfeatures{Scale=MatchLowercase}
  \defaultfontfeatures[\rmfamily]{Ligatures=TeX,Scale=1}
\fi
\IfFileExists{upquote.sty}{\usepackage{upquote}}{}
\IfFileExists{microtype.sty}{
  \usepackage[]{microtype}
  \UseMicrotypeSet[protrusion]{basicmath} 
}{}
\makeatletter
\@ifundefined{KOMAClassName}{
  \IfFileExists{parskip.sty}{%
    \usepackage{parskip}
  }{
    \setlength{\parindent}{0pt}
    \setlength{\parskip}{6pt plus 2pt minus 1pt}}
}{
  \KOMAoptions{parskip=half}}
\makeatother
\usepackage{xcolor}
\ifx\paragraph\undefined\else
  \let\oldparagraph\paragraph
  \renewcommand{\paragraph}[1]{\oldparagraph{#1}\mbox{}}
\fi
\ifx\subparagraph\undefined\else
  \let\oldsubparagraph\subparagraph
  \renewcommand{\subparagraph}[1]{\oldsubparagraph{#1}\mbox{}}
\fi

\usepackage{color}
\usepackage{fancyvrb}

\DefineVerbatimEnvironment{Highlighting}{Verbatim}{commandchars=\\\{\}}
\usepackage{framed}
\definecolor{shadecolor}{RGB}{241,243,245}
\newenvironment{Shaded}{\begin{snugshade}}{\end{snugshade}}

\newcommand{\AttributeTok}[1]{\textcolor[rgb]{0.40,0.45,0.13}{#1}}

\newcommand{\BuiltInTok}[1]{\textcolor[rgb]{0.00,0.23,0.31}{#1}}
\newcommand{\CharTok}[1]{\textcolor[rgb]{0.13,0.47,0.30}{#1}}
\newcommand{\CommentTok}[1]{\textcolor[rgb]{0.37,0.37,0.37}{#1}}

\newcommand{\ConstantTok}[1]{\textcolor[rgb]{0.56,0.35,0.01}{#1}}
\newcommand{\ControlFlowTok}[1]{\textcolor[rgb]{0.00,0.23,0.31}{#1}}
\newcommand{\DataTypeTok}[1]{\textcolor[rgb]{0.68,0.00,0.00}{#1}}
\newcommand{\DecValTok}[1]{\textcolor[rgb]{0.68,0.00,0.00}{#1}}
\newcommand{\DocumentationTok}[1]{\textcolor[rgb]{0.37,0.37,0.37}{\textit{#1}}}

\newcommand{\ExtensionTok}[1]{\textcolor[rgb]{0.00,0.23,0.31}{#1}}
\newcommand{\FloatTok}[1]{\textcolor[rgb]{0.68,0.00,0.00}{#1}}
\newcommand{\FunctionTok}[1]{\textcolor[rgb]{0.28,0.35,0.67}{#1}}

\newcommand{\KeywordTok}[1]{\textcolor[rgb]{0.00,0.23,0.31}{#1}}
\newcommand{\NormalTok}[1]{\textcolor[rgb]{0.00,0.23,0.31}{#1}}
\newcommand{\OperatorTok}[1]{\textcolor[rgb]{0.37,0.37,0.37}{#1}}
\newcommand{\OtherTok}[1]{\textcolor[rgb]{0.00,0.23,0.31}{#1}}

\newcommand{\SpecialCharTok}[1]{\textcolor[rgb]{0.37,0.37,0.37}{#1}}

\newcommand{\StringTok}[1]{\textcolor[rgb]{0.13,0.47,0.30}{#1}}
\newcommand{\VariableTok}[1]{\textcolor[rgb]{0.07,0.07,0.07}{#1}}

\providecommand{\tightlist}{%
  \setlength{\itemsep}{0pt}\setlength{\parskip}{0pt}}\usepackage{longtable,booktabs,array}
\usepackage{calc} 
\usepackage{etoolbox}
\makeatletter
\patchcmd\longtable{\par}{\if@noskipsec\mbox{}\fi\par}{}{}
\makeatother
\IfFileExists{footnotehyper.sty}{\usepackage{footnotehyper}}{\usepackage{footnote}}
\makesavenoteenv{longtable}
\usepackage{graphicx}
\makeatletter
\def\maxwidth{\ifdim\Gin@nat@width>\linewidth\linewidth\else\Gin@nat@width\fi}
\def\maxheight{\ifdim\Gin@nat@height>\textheight\textheight\else\Gin@nat@height\fi}
\makeatother
\setkeys{Gin}{width=\maxwidth,height=\maxheight,keepaspectratio}
\makeatletter
\def\fps@figure{htbp}
\makeatother
\newlength{\cslhangindent}
\newlength{\csllabelwidth}
\newlength{\cslentryspacingunit} 
\newenvironment{CSLReferences}[2] 
 {
  \setlength{\parindent}{0pt}
  \ifodd #1
  \let\oldpar\par
  \def\par{\hangindent=\cslhangindent\oldpar}
  \fi
  \setlength{\parskip}{#2\cslentryspacingunit}
 }%
 {}
\usepackage{calc}

\usepackage{arxiv}
\usepackage{orcidlink}
\usepackage{amsmath}
\usepackage[T1]{fontenc}
\makeatletter
\@ifpackageloaded{caption}{}{\usepackage{caption}}
\AtBeginDocument{%
\ifdefined\contentsname
  \renewcommand*\contentsname{Table of contents}
\else
  \newcommand\contentsname{Table of contents}
\fi
\ifdefined\listfigurename
  \renewcommand*\listfigurename{List of Figures}
\else
  \newcommand\listfigurename{List of Figures}
\fi
\ifdefined\listtablename
  \renewcommand*\listtablename{List of Tables}
\else
  \newcommand\listtablename{List of Tables}
\fi
\ifdefined\figurename
  \renewcommand*\figurename{Figure}
\else
  \newcommand\figurename{Figure}
\fi
\ifdefined\tablename
  \renewcommand*\tablename{Table}
\else
  \newcommand\tablename{Table}
\fi
}
\@ifpackageloaded{float}{}{\usepackage{float}}
\floatstyle{ruled}
\@ifundefined{c@chapter}{\newfloat{codelisting}{h}{lop}}{\newfloat{codelisting}{h}{lop}[chapter]}
\floatname{codelisting}{Listing}

\makeatother
\makeatletter
\@ifpackageloaded{caption}{}{\usepackage{caption}}
\@ifpackageloaded{subcaption}{}{\usepackage{subcaption}}
\makeatother
\makeatletter
\@ifpackageloaded{tcolorbox}{}{\usepackage[many]{tcolorbox}}
\makeatother
\makeatletter
\@ifundefined{shadecolor}{\definecolor{shadecolor}{rgb}{.97, .97, .97}}
\makeatother
\ifLuaTeX
  \usepackage{selnolig}  
\fi
\IfFileExists{bookmark.sty}{\usepackage{bookmark}}{\usepackage{hyperref}}
\IfFileExists{xurl.sty}{\usepackage{xurl}}{} 
\hypersetup{
  pdftitle={rang: Reconstructing reproducible R computational environments},
  pdfkeywords={R, reproducibility, docker},
  colorlinks=true,
  linkcolor={blue},
  filecolor={Maroon},
  citecolor={Blue},
  urlcolor={Blue},
  pdfcreator={LaTeX via pandoc}}

\newcommand{\runninghead}{A Preprint }
\title{rang: Reconstructing reproducible R computational environments}
\author{
\textbf{Chung-hong Chan}~\orcidlink{0000-0002-6232-7530}\\\\GESIS
Leibniz-Institut für Sozialwissenschaften\\\\\\\\\\
\textbf{David Schoch}~\orcidlink{0000-0003-2952-4812}\\\\GESIS
Leibniz-Institut für Sozialwissenschaften\\\\}
\date{}
\begin{document}
\maketitle
\begin{abstract}
A complete declarative description of the computational environment is
often missing when researchers share their materials. Without such
description, software obsolescence and missing system components can
jeopardize computational reproducibility in the future, even when data
and computer code are available. The R package rang is a complete
solution for generating the declarative description for other
researchers to automatically reconstruct the computational environment
at a specific time point. The reconstruction process, based on Docker,
has been tested for R code as old as 2001. The declarative description
generated by rang satisfies the definition of a reproducible research
compendium and can be shared as such. In this contribution, we show how
rang can be used to make otherwise unexecutable code, spanning from
fields such as computational social science and bioinformatics,
executable again. We also provide instructions on how to use rang to
construct reproducible and shareable research compendia of current
research. The package is currently available from CRAN
(https://cran.r-project.org/web/packages/rang/index.html) and GitHub
(https://github.com/chainsawriot/rang).
\end{abstract}
{\bfseries \emph Keywords}
\def\sep{\textbullet\ }
R \sep reproducibility \sep 
docker

\ifdefined\Shaded\renewenvironment{Shaded}{\begin{tcolorbox}[enhanced, interior hidden, boxrule=0pt, sharp corners, frame hidden, breakable, borderline west={3pt}{0pt}{shadecolor}]}{\end{tcolorbox}}\fi

\hypertarget{background}{%
\section{Background}\label{background}}

\emph{``In some cases the polarization estimation will not work \ldots{}
This is \emph{NOT} a problem in the method, it is entirely dependent on
the numpy version (and even the OS's). If you have different versions of
numpy or even the same version of numpy on a different OS configuration,
different networks will fail randomly\ldots{} {[}F{]}or instance, the
109th Congress will fail, but will work entirely normally on a different
numpy version, which will fail on a different Congress network.''}

\begin{quote}
\textasciitilde{} excerpt of
\href{https://www.michelecoscia.com/?page_id=2105}{this README file}
\end{quote}

Other than bad programming practices (Trisovic et al. 2022), the main
computing barrier to computational reproducibility is the failure to
reconstruct the computational environment like the one used by the
original researchers. This task looks trivially simple. But as computer
science research has shown, this task is incredibly complex (Abate et
al. 2015; Dolstra, Löh, and Pierron 2010). In the realm of a usual
scripting language such as R \footnote{In this paper, we will focus on
  R, a popular programming language used frequently in various
  computational fields (e.g.~computational social science,
  bioinformatics).}, that pertains four aspects: a) operating system, b)
system components such as \texttt{libxml2}, c) the exact R version, and
d) what and which version of the installed R packages. We will call them
Component A, B, C, D in the following sections. Any change in these four
components can possibly affect the execution of any shared computer
code. For example, the lack of the system component \texttt{libxml2} can
impact whether the R package \texttt{xml2} can be installed on a Linux
system. If the shared computer code requires the R package
\texttt{xml2}, then the whole execution fails.

In reality, the impact of Component A is relatively weak as mainstream,
open source programming languages and their software libraries are
usually cross platform. In modern computational research, Linux is the
de-facto operating system in high performance computing environments
(e.g.~Slurm). Instead, the impact of Components B, C, and D is much
higher. Component D is the most volatile among them all as there are
many possible combinations of R packages and versions. Software updates
with breaking changes (even in a dependency) might render existing
shared code using those changed features not executable or not producing
the same result anymore. Also, software obsolescence is commonplace,
especially since academic software is often not well maintained due to
lack of incentives (Merow et al. 2023).

The DevOps (software development and IT operations) community is also
confronted with this problem. The issue is usually half-jokingly
referred to as ``it works on my machine''-problem (Valstar, Griswold,
and Porter 2020, a software works on someone's local machine but is not
working anymore when deployed to the production system, indicates the
software tacitly depends on the computational environment of the local
machine). A partial solution to this problem from the DevOps community
is called \emph{containerization}. In essence, to containerize is to
develop and deploy the software together with all the libraries and the
operating system in an OS-level virtualization environment. In this way,
software dependency issues can be resolved inside the isolated
virtualized software environment and independent of what is installed on
the local computer. Docker is a popular choice in the DevOps world for
containerization.

To build a container, one needs to write a plain text declarative
description of the required computational environment. Inside this
declarative description, it should pin down all four Components
mentioned above. For Docker, it is in the form of a plain text file
called \texttt{Dockerfile}. This \texttt{Dockerfile} is then used as the
recipe to build a Docker image, where the four Components are assembled.
Then, one can launch a container with the built Docker image.

There has been many papers written on how containerization solutions
such as Docker can be helpful also to foster computational
reproducibility of science (e.g. Nüst and Hinz 2019; Peikert and
Brandmaier 2021; Boettiger and Eddelbuettel 2017). Although tutorials
are available (e.g. Nüst and Hinz 2019), providing a declarative
description of the computational environment in the form of Dockerfile
is far from the standard code sharing practice. This might be due to a
lack of (DevOps) skills of most scientists to create a Dockerfile (Kim,
Poline, and Dumas 2018). But there are many tools available to automate
the process (e.g. Nüst and Hinz 2019). The case in point described in
this paper, \texttt{rang}, is one of them. We argue that \texttt{rang}
is the only easy-to-use solution available that can pin down and restore
all four components without the reliance on any commercial service such
as MRAN.

\hypertarget{existing-solutions}{%
\subsection{Existing solutions}\label{existing-solutions}}

\texttt{renv} (Ushey 2022) (and its derivatives such as \texttt{jetpack}
and its predecessor \texttt{packrat}) takes a similar approach to
Python's \texttt{virtualenv} and Ruby's \texttt{Gem} to pin down the
exact version of R packages using a ``lock file''. Other solutions such
as \texttt{checkpoint} (Ooi, de Vries, and Microsoft 2022) depend on the
availability of The Microsoft R Application Network (MRAN, a
time-stamped daily backup of CRAN), which will be shut down on July 1st,
2023. \texttt{groundhog} (Simonsohn and Gruson 2023) used to depend on
MRAN but has a plan to switch to their home-grown R package repository.
These solution can effectively pin down Component C and D. But they can
only restore component D. Also, for solutions depending on MRAN, there
is a limit on how far back this reproducibility can go, since MRAN can
only go back as far as September 17, 2014. Additionally, it only covers
CRAN packages.

\texttt{containerit} (Nüst and Hinz 2019) takes the current state of the
computational environment and documents it as a Dockerfile.
\texttt{containerit} makes the assumption that Component A has a weak
influence on computational reproducibility and therefore defaults to
Linux-based Rocker images. In this way, it fixes Component A. But
\texttt{containerit} does not pin down the exact version of R packages.
Therefore, it can pin down components A, B, C, but only a part of
component D. \texttt{dockta} is another containerization solution that
can potentially pin down all components due to the fact that MRAN is
used. But it also suffers from the same limitations mentioned above.

It is also worth mentioning that MRAN is not the only archival service.
Posit also provides a free (\emph{gratis}) time-stamped daily backup of
CRAN and Bioconductor (a series of repositories of R package for
bioinformatics and computational biology) called Posit Public Package
Manager
(https://packagemanager.rstudio.com/client/\#/repos/2/packages/). It can
goes as far back as October 10, 2017.

These solutions are better for prospective usage, i.e.~using them now to
ensure the reproducibility of the current research for future
researchers. \texttt{rang} mostly targets retrospective usage,
i.e.~using \texttt{rang} to reconstruct historical R computational
environments for which the declarative descriptions are not available.
One can think of \texttt{rang} as an archaeological tool. In this realm,
we could not find any existing solution targeting R specifically which
does not currently depend on MRAN.

\hypertarget{structure-of-this-paper}{%
\subsection{Structure of this paper}\label{structure-of-this-paper}}

In Section 2, we will explain how to use \texttt{rang}. In Section 3,
\texttt{rang} is used to enable the reproducibility of published
literature (with increasing sophistication). However, one can still use
\texttt{rang} for prospective usage and arguably can ensure a longer
term computational reproducibility than other solutions. In Section 4,
\texttt{rang} is used to create an executable research compendium with
Docker and \texttt{Make} (Baker 2020).

\hypertarget{basic-usage}{%
\section{Basic usage}\label{basic-usage}}

There are two important functions of \texttt{rang}: \texttt{resolve()}
and \texttt{dockerize()}.

\texttt{resolve()} queries various web services from the r-hub project
of the R Consortium for information about R packages at a specific time
point that is necessary for reconstructing a computational environment,
e.g.~(deep) dependencies (Component D), R version (Component C), and
system requirements (Component B). For instance, if there was a
computational environment constructed on 2020-01-16 (called ``snapshot
date'') with the several natural language processing R packages,
\texttt{resolve()} can be used to resolve all the dependencies of these
R packages. Currently, \texttt{rang} supports CRAN, Bioconductor,
GitHub, and local packages.

\begin{Shaded}
\begin{Highlighting}[]
\FunctionTok{library}\NormalTok{(rang)}
\NormalTok{graph }\OtherTok{\textless{}{-}} \FunctionTok{resolve}\NormalTok{(}\AttributeTok{pkgs =} \FunctionTok{c}\NormalTok{(}\StringTok{"openNLP"}\NormalTok{, }\StringTok{"LDAvis"}\NormalTok{, }\StringTok{"topicmodels"}\NormalTok{, }\StringTok{"quanteda"}\NormalTok{),}
                 \AttributeTok{snapshot\_date =} \StringTok{"2020{-}01{-}16"}\NormalTok{)}
\NormalTok{graph}
\end{Highlighting}
\end{Shaded}

The resolved result is an S3 object called \texttt{rang}. The
information contained in a \texttt{rang} object can then be used to
construct a computational environment in a similar manner as
\texttt{containerit}, but with the packages and R versions pinned on the
snapshot date. Then, the function \texttt{dockerize()} is used to
generate the \texttt{Dockerfile} and other scripts in the
\texttt{output\_dir}.

\begin{Shaded}
\begin{Highlighting}[]
\FunctionTok{dockerize}\NormalTok{(graph, }\AttributeTok{output\_dir =} \StringTok{"docker"}\NormalTok{)}
\end{Highlighting}
\end{Shaded}

For R \textgreater= 3.1, the images from the Rocker project are used
(Boettiger and Eddelbuettel 2017). For R \textless{} 3.1 but
\textgreater= 1.3.1, a custom image based on Debian is used. As of
writing, \texttt{rang} does not support R \textless{} 1.3.1,
i.e.~snapshot date earlier than 2001-08-31 (which is 13 years earlier
than all solutions depending on MRAN). There are two features of
\texttt{dockerize()} that are important for future reproducibility.

\begin{enumerate}
\def\labelenumi{\arabic{enumi}.}
\tightlist
\item
  By default, the container building process downloads source packages
  from their sources and then compiles them. This step depends on the
  future availability of R packages on CRAN (which is extremely likely
  to be the case in the near future, given the continuous availability
  since 1997-04-23) \footnote{https://stat.ethz.ch/pipermail/r-announce/1997/000001.html},
  Bioconductor, and Github. However, it is also possible to cache (or
  archive) the source packages now. The archived R packages can then be
  used instead during the building process. The significance of this
  step in terms of long-term computational reproducibility will be
  discussed in Section 4.
\end{enumerate}

\begin{Shaded}
\begin{Highlighting}[]
\FunctionTok{dockerize}\NormalTok{(graph, }\AttributeTok{output\_dir =} \StringTok{"docker"}\NormalTok{, }\AttributeTok{cache =} \ConstantTok{TRUE}\NormalTok{)}
\end{Highlighting}
\end{Shaded}

\begin{enumerate}
\def\labelenumi{\arabic{enumi}.}
\setcounter{enumi}{1}
\tightlist
\item
  It is also possible to install R packages in a separate library during
  the building process to isolate all these R packages from the main
  library.
\end{enumerate}

\begin{Shaded}
\begin{Highlighting}[]
\FunctionTok{dockerize}\NormalTok{(graph, }\AttributeTok{output\_dir =} \StringTok{"docker"}\NormalTok{, }\AttributeTok{cache =} \ConstantTok{TRUE}\NormalTok{,}
          \AttributeTok{lib =} \StringTok{"anotherlibrary"}\NormalTok{)}
\end{Highlighting}
\end{Shaded}

For the sake of completeness, the instructions for building and running
the Docker container on Unix-like systems are included here.

\begin{Shaded}
\begin{Highlighting}[]
\BuiltInTok{cd}\NormalTok{ docker}
\CommentTok{\#\# might need to sudo}
\ExtensionTok{docker}\NormalTok{ build }\AttributeTok{{-}t}\NormalTok{ rang .}
\CommentTok{\#\# interactive environment}
\ExtensionTok{docker}\NormalTok{ run }\AttributeTok{{-}{-}rm} \AttributeTok{{-}{-}name} \StringTok{"rangtest"} \AttributeTok{{-}ti}\NormalTok{ rang}
\end{Highlighting}
\end{Shaded}

\hypertarget{project-scanning}{%
\subsection{Project scanning}\label{project-scanning}}

The first argument of \texttt{resolve()} is processed by a separate
function called \texttt{as\_pkgrefs()}. For interoperability,
\texttt{rang} supports the ``package references'' standard \footnote{https://r-lib.github.io/pkgdepends/reference/pkg\_refs.html}
used also in other packages such as \texttt{renv} (Ushey 2022). It is
mostly used for converting ``shorthands'' (e.g.~\texttt{xml2} and
\texttt{S4Vectors}) to package references (e.g.~\texttt{cran::xml2} and
\texttt{bioc::S4Vectors}).

When \texttt{as\_pkgrefs()} is applied to a single path of a directory,
it scans all relevant files (\texttt{DESCRIPTION}, R scripts and R
Markdown files) for all R packages used (based on
\texttt{renv::dependencies()} ). How it works is demonstrated in three
of the following examples below. But an important caveat is that it can
only scan CRAN and Bioconductor packages.

\hypertarget{case-studies}{%
\section{Case Studies}\label{case-studies}}

The following are some examples of how \texttt{rang} can be used to make
shared, but otherwise unexecutable, R code runnable again. The examples
were drawn from various fields spanning from political science,
psychological science, and bioinformatics.

\hypertarget{quanteda-joss-paper}{%
\subsection{quanteda JOSS paper}\label{quanteda-joss-paper}}

The software paper of the text analysis R package \texttt{quanteda} was
published on 2018-10-06 (Benoit et al. 2018). In the paper, the
following R code snippet is included.

\begin{Shaded}
\begin{Highlighting}[]
\FunctionTok{library}\NormalTok{(}\StringTok{"quanteda"}\NormalTok{)}
\CommentTok{\# construct the feature co{-}occurrence matrix}
\NormalTok{examplefcm }\OtherTok{\textless{}{-}}
\FunctionTok{tokens}\NormalTok{(data\_corpus\_irishbudget2010, }\AttributeTok{remove\_punct =} \ConstantTok{TRUE}\NormalTok{) }\SpecialCharTok{\%\textgreater{}\%}
\FunctionTok{tokens\_tolower}\NormalTok{() }\SpecialCharTok{\%\textgreater{}\%}
\FunctionTok{tokens\_remove}\NormalTok{(}\FunctionTok{stopwords}\NormalTok{(}\StringTok{"english"}\NormalTok{), }\AttributeTok{padding =} \ConstantTok{FALSE}\NormalTok{) }\SpecialCharTok{\%\textgreater{}\%}
\FunctionTok{fcm}\NormalTok{(}\AttributeTok{context =} \StringTok{"window"}\NormalTok{, }\AttributeTok{window =} \DecValTok{5}\NormalTok{, }\AttributeTok{tri =} \ConstantTok{FALSE}\NormalTok{)}
\CommentTok{\# choose 30 most frequency features}
\NormalTok{topfeats }\OtherTok{\textless{}{-}} \FunctionTok{names}\NormalTok{(}\FunctionTok{topfeatures}\NormalTok{(examplefcm, }\DecValTok{30}\NormalTok{))}
\CommentTok{\# select the top 30 features only, plot the network}
\FunctionTok{set.seed}\NormalTok{(}\DecValTok{100}\NormalTok{)}
\FunctionTok{textplot\_network}\NormalTok{(}\FunctionTok{fcm\_select}\NormalTok{(examplefcm, topfeats), }\AttributeTok{min\_freq =} \FloatTok{0.8}\NormalTok{)}
\end{Highlighting}
\end{Shaded}

On 2023-02-08, this code snippet is not executable with the current
version of \texttt{quanteda} (3.2.4). It is possible to install the
``period appropriate'' version of \texttt{quanteda} (1.3.4) using
\texttt{remotes} on the current version of R (4.2.2). And indeed, the
above code snippet can still be executed.

\begin{Shaded}
\begin{Highlighting}[]
\NormalTok{remotes}\SpecialCharTok{::}\FunctionTok{install\_version}\NormalTok{(}\StringTok{"quanteda"}\NormalTok{, }\AttributeTok{version =} \StringTok{"1.3.4"}\NormalTok{)}
\end{Highlighting}
\end{Shaded}

The issue is that installing \texttt{quanteda} 1.3.4 this way installs
the latest dependencies from CRAN. \texttt{quanteda} 1.3.4 uses a
deprecated (but not yet removed) function of \texttt{Matrix}
(\texttt{as(\textless{}dgTMatrix\textgreater{},\ "dgCMatrix")}). If this
function were removed in the future, the above code snippet would not be
executable anymore.

Using \texttt{rang}, one can query the version of \texttt{quanteda} on
2018-10-06 and create a Docker container with all the ``period
appropriate'' dependencies. Here, the \texttt{rstudio} Rocker image is
selected.

\begin{Shaded}
\begin{Highlighting}[]
\FunctionTok{library}\NormalTok{(rang)}
\NormalTok{graph }\OtherTok{\textless{}{-}} \FunctionTok{resolve}\NormalTok{(}\AttributeTok{pkgs =} \StringTok{"quanteda"}\NormalTok{,}
                 \AttributeTok{snapshot\_date =} \StringTok{"2018{-}10{-}06"}\NormalTok{,}
                 \AttributeTok{os =} \StringTok{"ubuntu{-}18.04"}\NormalTok{)}
\FunctionTok{dockerize}\NormalTok{(graph, }\AttributeTok{output\_dir =} \StringTok{"quanteda\_docker"}\NormalTok{,}
          \AttributeTok{image =} \StringTok{"rstudio"}\NormalTok{)}
\end{Highlighting}
\end{Shaded}

\begin{figure}

{\centering \includegraphics[width=4in,height=\textheight]{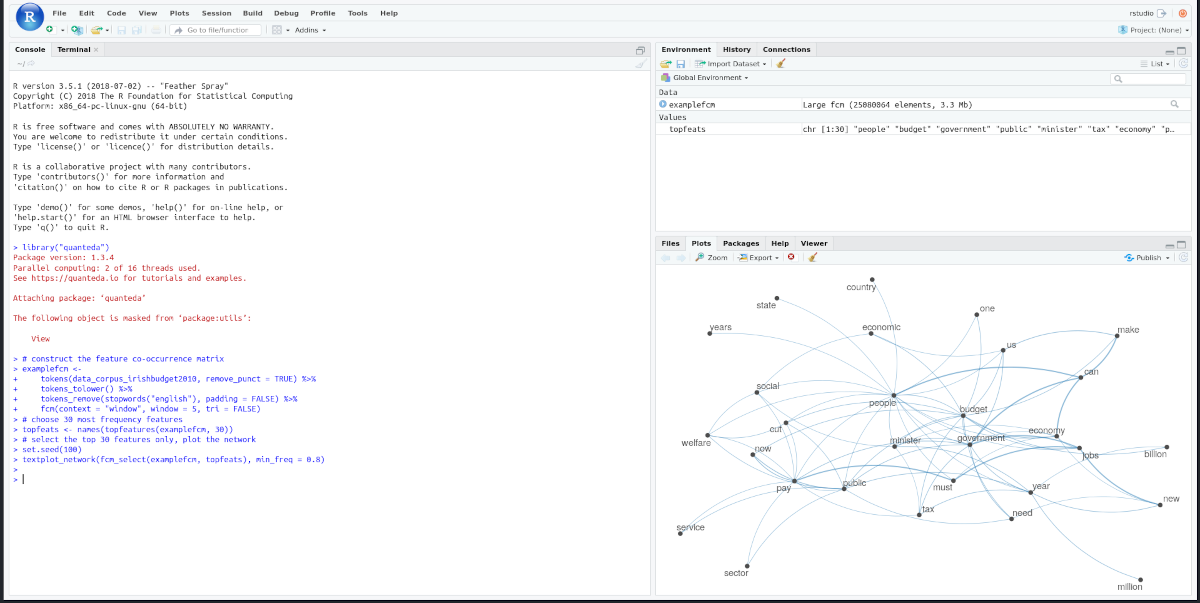}

}

\caption{\label{fig-figure1}The code snippet running in a R 3.5.1
container created with rang}

\end{figure}

The above code snippet can be executed with the generated container
without any problem (Figure~\ref{fig-figure1}).

\hypertarget{psychological-science}{%
\subsection{Psychological Science}\label{psychological-science}}

Crüwell et al. (2023) evaluate the computational reproducibility of 14
articles published in \emph{Psyhocological Science}. Among these
articles, the paper by Hilgard et al. (2019) has been rated as having
``package dependency issues''.

All data and computer code are available from GitHub with the last
commit on 2019-01-17 \footnote{https://github.com/Joe-Hilgard/vvg-2d4d}.
The R code contains a list of R packages used in the project as
\texttt{library()} statements, including an R package on GitHub that is
written by the main author of that paper. However, we identified one
package (\texttt{compute.es}) that was not written in those
\texttt{library()} statements but used with the namespace operator,
i.e.~\texttt{compute.es::tes()}. This undocumented package can be
detected by \texttt{renv::dependencies()}, which is the provider of the
scanning function of \texttt{rang}.

Based on the above information, one can run \texttt{resolve()} to obtain
the dependency graph of all R packages on 2019-01-17.

\begin{Shaded}
\begin{Highlighting}[]
\DocumentationTok{\#\# scan all packages}
\NormalTok{r\_pkgs }\OtherTok{\textless{}{-}} \FunctionTok{as\_pkgrefs}\NormalTok{(}\StringTok{"vvg{-}2d4d"}\NormalTok{)}
\DocumentationTok{\#\# replace cran::hilgard with Github}
\NormalTok{r\_pkgs[r\_pkgs }\SpecialCharTok{==} \StringTok{"cran::hilhard"}\NormalTok{] }\OtherTok{\textless{}{-}} \StringTok{"Joe{-}Hilgard/hilgard"}
\NormalTok{graph }\OtherTok{\textless{}{-}} \FunctionTok{resolve}\NormalTok{(r\_pkgs, }\AttributeTok{snapshot\_date =} \StringTok{"2019{-}01{-}17"}\NormalTok{)}
\end{Highlighting}
\end{Shaded}

When running \texttt{dockerize()}, one can take advantage of the
\texttt{materials\_dir} parameter to transfer the shared materials from
Hilgard et al. (2019) into the Docker image.

\begin{Shaded}
\begin{Highlighting}[]
\FunctionTok{dockerize}\NormalTok{(graph, }\StringTok{"hilgard"}\NormalTok{, }\AttributeTok{materials\_dir =} \StringTok{"vvg{-}2d4d"}\NormalTok{, }\AttributeTok{cache =} \ConstantTok{TRUE}\NormalTok{)}
\end{Highlighting}
\end{Shaded}

We then built the Docker and launch a Docker container. For this
container, we changed the entry point from R to bash so that the
container goes to the Linux command shell instead.

\begin{Shaded}
\begin{Highlighting}[]
\NormalTok{cd hilgard}
\NormalTok{docker build }\SpecialCharTok{{-}}\NormalTok{t hilgard .}
\NormalTok{docker run }\SpecialCharTok{{-}{-}}\NormalTok{rm }\SpecialCharTok{{-}{-}}\NormalTok{name }\StringTok{"hilgardcontainer"} \SpecialCharTok{{-}{-}}\NormalTok{entrypoint bash }\SpecialCharTok{{-}}\NormalTok{ti hilgard}
\end{Highlighting}
\end{Shaded}

Inside the container, the materials are located in the
\texttt{materials} directory. We used the following shell script to test
the reproducibility of all R scripts.

\begin{Shaded}
\begin{Highlighting}[]
\BuiltInTok{cd}\NormalTok{ materials}
\VariableTok{rfiles}\OperatorTok{=}\VariableTok{(}\NormalTok{0\_data\_aggregation.R 1\_data\_cleaning.R 2\_analysis.R 3\_plotting.R}\VariableTok{)}
\ControlFlowTok{for}\NormalTok{ i }\KeywordTok{in} \VariableTok{$\{rfiles}\OperatorTok{[@]}\VariableTok{\}}
\ControlFlowTok{do}
    \ExtensionTok{Rscript} \VariableTok{$i}
    \VariableTok{code}\OperatorTok{=}\VariableTok{$?}
    \ControlFlowTok{if} \BuiltInTok{[} \VariableTok{$code} \OtherTok{!=}\NormalTok{ 0 }\BuiltInTok{]}
    \ControlFlowTok{then}
        \BuiltInTok{exit}\NormalTok{ 1}
    \ControlFlowTok{fi}
\ControlFlowTok{done}
\end{Highlighting}
\end{Shaded}

All R scripts ran fine inside the container and the figures generated
are the same as the ones in Hilgard et al. (2019).

\hypertarget{political-analysis}{%
\subsection{Political Analysis}\label{political-analysis}}

The study by Trisovic et al. (2022) evaluates the reproducibility of R
scripts shared on Dataverse. They found that 75\% of R scripts cannot be
successfully executed. Among these failed R scripts is an R script
shared by Beck (2019).

This R script has been ``rescued'' by the author of the R package
\texttt{groundhog} (Simonsohn and Gruson 2023), as demonstrated in a
blog post \footnote{http://datacolada.org/100}. We were wondering if
\texttt{rang} can also be used to ``rescue'' the concerned R script. The
date of the R script, as indicated on Dataverse, is 2018-12-12. This
date is used as the snapshot date.

\begin{Shaded}
\begin{Highlighting}[]
\DocumentationTok{\#\# as\_pkgrefs is automatically run in this case}
\NormalTok{graph }\OtherTok{\textless{}{-}} \FunctionTok{resolve}\NormalTok{(}\StringTok{"nathaniel"}\NormalTok{, }\AttributeTok{snapshot\_date =} \StringTok{"2018{-}12{-}12"}\NormalTok{)}
\FunctionTok{dockerize}\NormalTok{(graph, }\AttributeTok{output\_dir =} \StringTok{"nat"}\NormalTok{, }\AttributeTok{materials\_dir =} \StringTok{"nathaniel"}\NormalTok{)}
\end{Highlighting}
\end{Shaded}

\begin{Shaded}
\begin{Highlighting}[]
\NormalTok{cd nat}
\NormalTok{docker build }\SpecialCharTok{{-}}\NormalTok{t nat .}
\NormalTok{docker run }\SpecialCharTok{{-}{-}}\NormalTok{rm }\SpecialCharTok{{-}{-}}\NormalTok{name }\StringTok{"natcontainer"} \SpecialCharTok{{-}{-}}\NormalTok{entrypoint bash }\SpecialCharTok{{-}}\NormalTok{ti nat}
\end{Highlighting}
\end{Shaded}

Inside the container

\begin{Shaded}
\begin{Highlighting}[]
\NormalTok{cd materials}
\NormalTok{Rscript fn\_5.R}
\end{Highlighting}
\end{Shaded}

The same file can thus also be ``rescued'' by \texttt{rang}.

\hypertarget{recover-a-removed-r-package-maxent}{%
\subsection{Recover a removed R package:
maxent}\label{recover-a-removed-r-package-maxent}}

The R package \texttt{maxent} introduces a machine learning algorithm
with a small memory footprint and was available on CRAN until 2019. A
software paper was published by the original authors in 2012 (Jurka
2012). The R package was also used in some subsequent automated content
analytic papers (e.g. Lörcher and Taddicken 2017). Despite the covert
editing of the package by a staffer of CRAN \footnote{https://github.com/cran/maxent/commit/9d46c6aad27a1f41a78907b170ddd9a586192be9},
the package was removed from CRAN in 2019 \footnote{https://cran-archive.r-project.org/web/checks/2019/2019-03-05\_check\_results\_maxent.html}.
We attempted to install the second last (the original submitted version)
and last (with covert editing) versions of \texttt{maxent} on R 4.2.2.
Both of them didn't work.

Using \texttt{rang}, we are able to reconstruct a computational
environment with R 2.15.0 (2012-03-30) to run all code snippets
published in Jurka (2012) \footnote{On an interesting historical side
  note: The original paper reported ---based on a benchmark--- that
  ``the algorithm is very fast; \texttt{maxent} uses only 135.4
  megabytes of RAM and finishes in 53.3 seconds.'' On a modest computer
  in 2023 with a dockerized R 2.15.0, the benchmark finishes in 4
  seconds.}. For removed CRAN packages, we strongly recommend querying
the Github read-only mirror of CRAN instead (https://github.com/cran).
It is because in this way, the resolved system requirements have a
higher chance of being correct.

\begin{Shaded}
\begin{Highlighting}[]
\NormalTok{maxent }\OtherTok{\textless{}{-}} \FunctionTok{resolve}\NormalTok{(}\StringTok{"cran/maxent"}\NormalTok{, }\StringTok{"2012{-}06{-}10"}\NormalTok{)}
\FunctionTok{dockerize}\NormalTok{(maxent, }\StringTok{"maxentdir"}\NormalTok{, }\AttributeTok{cache =} \ConstantTok{TRUE}\NormalTok{)}
\end{Highlighting}
\end{Shaded}

\hypertarget{recover-a-removed-r-package-ptproc}{%
\subsection{Recover a removed R package:
ptproc}\label{recover-a-removed-r-package-ptproc}}

The software paper of the R package \texttt{ptproc} was published in
2003 and introduced multidimensional point process models (Peng 2003).
But the package has been removed from CRAN for over a decade (at least).
The only release on CRAN was on 2002-10-10. The package is still listed
in the ``Handling and Analyzing Spatio-Temporal Data'' CRAN Task View
\footnote{https://cran.r-project.org/web/views/SpatioTemporal.html}
despite being uninstallable without modification on any modern R system
(see below). As of writing, the package, as a tarball file (tar.gz), is
still downloadable from the original author's website \footnote{https://www.biostat.jhsph.edu/\textasciitilde rpeng/software/}.

Even with this over-a-decade removal and new packages with similar
functionalities have been created, there is evidence that
\texttt{ptproc} is still being sought for. As late as 2017, there are
blog posts on how to install the long obsolete package on modern
versions of R \footnote{https://blog.mathandpencil.com/installing-ptproc-on-osx
  and
  https://tomaxent.com/2017/03/16/Installing-ptproc-on-Ubuntu-16-04-LTS/}.
The package is extremely challenging to install on a modern R system
because the package was written before the introduction of name space
management in R 1.7.0 (Tierney 2003). In other words, the available
tarball file from the original author's website does not contain a
\texttt{NAMESPACE} file as all other modern R packages do.

The oldest version of R that \texttt{rang} can support, as of writing,
is R 1.3.1. \texttt{rang} is probably the only solution available that
can support the 1.x series of R (i.e.~before 2004-10-04). Similar to the
case of \texttt{maxent} above, a Dockerfile to assemble a Docker image
with \texttt{ptproc} installed can be generated with two lines of code.

\begin{Shaded}
\begin{Highlighting}[]
\NormalTok{graph }\OtherTok{\textless{}{-}} \FunctionTok{resolve}\NormalTok{(}\StringTok{"ptproc"}\NormalTok{, }\AttributeTok{snapshot\_date =} \StringTok{"2004{-}07{-}01"}\NormalTok{)}
\FunctionTok{dockerize}\NormalTok{(graph, }\StringTok{"\textasciitilde{}/dev/misc/ptproc"}\NormalTok{, }\AttributeTok{cache =} \ConstantTok{TRUE}\NormalTok{)}
\end{Highlighting}
\end{Shaded}

Suppose we have an R script, extracted from Peng (2003), called
``peng.R'' like this:

\begin{Shaded}
\begin{Highlighting}[]
\FunctionTok{require}\NormalTok{(ptproc)}

\FunctionTok{set.seed}\NormalTok{(}\DecValTok{1000}\NormalTok{)}
\NormalTok{x }\OtherTok{\textless{}{-}} \FunctionTok{cbind}\NormalTok{(}\FunctionTok{runif}\NormalTok{(}\DecValTok{100}\NormalTok{), }\FunctionTok{runif}\NormalTok{(}\DecValTok{100}\NormalTok{), }\FunctionTok{runif}\NormalTok{(}\DecValTok{100}\NormalTok{))}
\NormalTok{hPois.cond.int }\OtherTok{\textless{}{-}} \ControlFlowTok{function}\NormalTok{(params, eval.pts, }\AttributeTok{pts =} \ConstantTok{NULL}\NormalTok{, }\AttributeTok{data =} \ConstantTok{NULL}\NormalTok{, }\AttributeTok{TT =} \ConstantTok{NULL}\NormalTok{) \{}
\NormalTok{    mu }\OtherTok{\textless{}{-}}\NormalTok{ params[}\DecValTok{1}\NormalTok{]}
    \ControlFlowTok{if}\NormalTok{(}\FunctionTok{is.null}\NormalTok{(TT))}
        \FunctionTok{rep}\NormalTok{(mu, }\FunctionTok{nrow}\NormalTok{(eval.pts))}
    \ControlFlowTok{else}\NormalTok{ \{}
\NormalTok{        vol }\OtherTok{\textless{}{-}} \FunctionTok{prod}\NormalTok{(}\FunctionTok{apply}\NormalTok{(TT, }\DecValTok{2}\NormalTok{, diff))}
\NormalTok{        mu }\SpecialCharTok{*}\NormalTok{ vol}
\NormalTok{    \}}
\NormalTok{\}}
\NormalTok{ppm }\OtherTok{\textless{}{-}} \FunctionTok{ptproc}\NormalTok{(}\AttributeTok{pts =}\NormalTok{ x, }\AttributeTok{cond.int =}\NormalTok{ hPois.cond.int, }\AttributeTok{params =} \DecValTok{50}\NormalTok{,}
              \AttributeTok{ranges =} \FunctionTok{cbind}\NormalTok{(}\FunctionTok{c}\NormalTok{(}\DecValTok{0}\NormalTok{,}\DecValTok{1}\NormalTok{), }\FunctionTok{c}\NormalTok{(}\DecValTok{0}\NormalTok{,}\DecValTok{1}\NormalTok{), }\FunctionTok{c}\NormalTok{(}\DecValTok{0}\NormalTok{,}\DecValTok{1}\NormalTok{)))}
\NormalTok{fit }\OtherTok{\textless{}{-}} \FunctionTok{ptproc.fit}\NormalTok{(ppm, }\AttributeTok{optim.control =} \FunctionTok{list}\NormalTok{(}\AttributeTok{trace =} \DecValTok{2}\NormalTok{), }\AttributeTok{method =} \StringTok{"BFGS"}\NormalTok{)}
\FunctionTok{summary}\NormalTok{(fit)}
\end{Highlighting}
\end{Shaded}

One can integrate \texttt{rang} into a BASH script to completely
automate the batch execution of the above R script.

\begin{Shaded}
\begin{Highlighting}[]
\ExtensionTok{Rscript} \AttributeTok{{-}e} \StringTok{"require(rang); dockerize(resolve(\textquotesingle{}ptproc\textquotesingle{}, \textquotesingle{}2004{-}07{-}01\textquotesingle{}),}
\StringTok{\textquotesingle{}pengdocker\textquotesingle{}, cache = TRUE)"}
\ExtensionTok{docker}\NormalTok{ build }\AttributeTok{{-}t}\NormalTok{ pengimg ./pengdocker}
\CommentTok{\#\# launching a container in daemon mode {-}d}
\ExtensionTok{docker}\NormalTok{ run }\AttributeTok{{-}d} \AttributeTok{{-}{-}rm} \AttributeTok{{-}{-}name} \StringTok{"pengcontainer"} \AttributeTok{{-}ti}\NormalTok{ pengimg}
\ExtensionTok{docker}\NormalTok{ cp peng.R pengcontainer:/peng.R}
\ExtensionTok{docker}\NormalTok{ exec pengcontainer R CMD BATCH peng.R}
\ExtensionTok{docker}\NormalTok{ exec pengcontainer cat peng.Rout}
\ExtensionTok{docker}\NormalTok{ cp pengcontainer:/peng.Rout peng.Rout}
\ExtensionTok{docker}\NormalTok{ stop pengcontainer}
\end{Highlighting}
\end{Shaded}

The file \texttt{peng.Rout} contains the execution results of the script
from inside the Docker container. As the random seed was preserved by
the original author (Peng 2003), the above BASH script can perfectly
reproduce the analysis \footnote{It is also important to note that the
  random number generator (RNG) of R has been changed several times over
  the course of the development. In this case, we are using the same
  generation of RNG as Peng (2003).}.

\hypertarget{recover-a-removed-bioconductor-package}{%
\subsection{Recover a removed Bioconductor
package}\label{recover-a-removed-bioconductor-package}}

Similar to CRAN, packages can also be removed over time from
Bioconductor. The Bioconductor package \texttt{Sushi} has been
deprecated by the original authors and is removed from Bioconductor
version 3.16 (2022-11-02). \texttt{Sushi} is a data visualization tool
for genomic data and was used in many online tutorials and scientific
papers, including the original paper announcing the package by the
original authors (Phanstiel et al. 2014).

\begin{figure}

{\centering \includegraphics{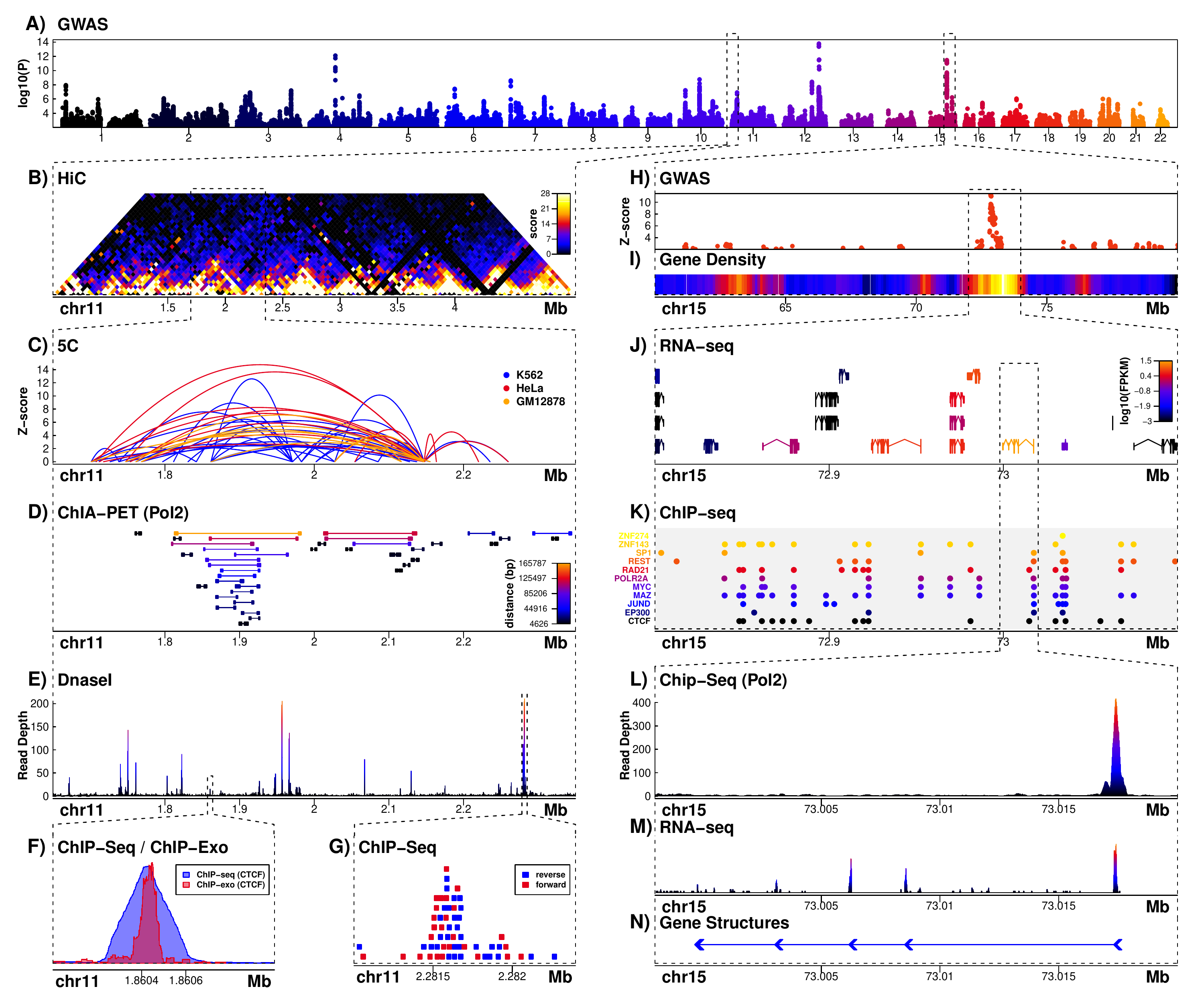}

}

\caption{\label{fig-figure2}The figure from the batch execution of
\texttt{PaperFigure.R} inside a Docker container generated by
\texttt{rang}}

\end{figure}

\texttt{rang} has native support for Bioconductor packages since version
0.2. We obtained the R script \texttt{"PaperFigure.R"} from the Github
repository of \texttt{Sushi} \footnote{https://github.com/PhanstielLab/Sushi/blob/master/vignettes/PaperFigure.R},
which generates the figure in Phanstiel et al. (2014). Similar to the
above case of \texttt{ptproc}, we made a completely automated BASH
script to run \texttt{"PaperFigure.R"} and get the generated figure out
of the container (Figure~\ref{fig-figure2}). We made no modification to
\texttt{"PaperFigure.R"}.

\begin{Shaded}
\begin{Highlighting}[]
\ExtensionTok{Rscript} \AttributeTok{{-}e} \StringTok{"require(rang); dockerize(resolve(\textquotesingle{}Sushi\textquotesingle{}, \textquotesingle{}2014{-}06{-}05\textquotesingle{}),}
\StringTok{\textquotesingle{}sushidocker\textquotesingle{}, no\_rocker = TRUE, cache = TRUE)"}
\ExtensionTok{docker}\NormalTok{ build }\AttributeTok{{-}t}\NormalTok{ sushiimg ./sushidocker}
\ExtensionTok{docker}\NormalTok{ run }\AttributeTok{{-}d} \AttributeTok{{-}{-}rm} \AttributeTok{{-}{-}name} \StringTok{"sushicontainer"} \AttributeTok{{-}ti}\NormalTok{ sushiimg}
\ExtensionTok{docker}\NormalTok{ cp PaperFigure.R sushicontainer:/PaperFigure.R}
\ExtensionTok{docker}\NormalTok{ exec sushicontainer mkdir vignettes}
\ExtensionTok{docker}\NormalTok{ exec sushicontainer R CMD BATCH PaperFigure.R}
\ExtensionTok{docker}\NormalTok{ cp sushicontainer:/vignettes/Figure\_1.pdf sushi\_figure1.pdf}
\ExtensionTok{docker}\NormalTok{ stop sushicontainer}
\end{Highlighting}
\end{Shaded}

\hypertarget{preparing-research-compendia-with-long-term-computational-reproducibility}{%
\section{Preparing research compendia with long-term computational
reproducibility}\label{preparing-research-compendia-with-long-term-computational-reproducibility}}

The above six examples show how powerful \texttt{rang} is to reconstruct
tricky computational environments which have not been completely
declared in the literature. Although we position \texttt{rang} mostly as
an archaeological tool, we think that \texttt{rang} can also be used to
prepare research compendia of current research. We can't predict the
future but research compendia generated by \texttt{rang} would probably
have long-term computational reproducibility.

To demonstrate this point, we took the recent paper by Oser et al.
(2022). This paper was selected because 1) the paper was published in
\emph{Political Communication}, a high impact journal that awards Open
Science Badges; 2) shared data and R code are available; and most
importantly, 3) the shared R code is well-written. In the repository of
this paper, we based on the materials shared by Oser et al. (2022) and
prepared a research compendium that should have long-term computational
reproducibility. The research compendium is similar to the Executable
Compendium suggested by the Turing way.

The preparation of the research compendium is easy as \texttt{rang} can
scan a materials directory for all R packages used \footnote{We detected
  a minor issue in the code base that an undeclared Github package is
  used. But it can be easily solved, as in the Psychological Science
  example above.}.

\begin{Shaded}
\begin{Highlighting}[]
\FunctionTok{require}\NormalTok{(rang)}
\DocumentationTok{\#\# meta{-}analysis is the directory of all shared materials}
\NormalTok{cran\_pkgs }\OtherTok{\textless{}{-}} \FunctionTok{as\_pkgrefs}\NormalTok{(}\StringTok{"meta{-}analysis"}\NormalTok{) }

\DocumentationTok{\#\# dmetar is an undeclared github package: MathiasHarrer/dmetar}
\NormalTok{cran\_pkgs[cran\_pkgs }\SpecialCharTok{==} \StringTok{"cran::dmetar"}\NormalTok{] }\OtherTok{\textless{}{-}} \StringTok{"MathiasHarrer/dmetar"}
\NormalTok{x }\OtherTok{\textless{}{-}} \FunctionTok{resolve}\NormalTok{(cran\_pkgs, }\StringTok{"2021{-}08{-}11"}\NormalTok{, }\AttributeTok{verbose =} \ConstantTok{TRUE}\NormalTok{)}
\DocumentationTok{\#\#print(x, all\_pkgs = TRUE)}
\FunctionTok{dockerize}\NormalTok{(x, }\StringTok{"oserdocker"}\NormalTok{, }\AttributeTok{materials\_dir =} \StringTok{"meta{-}analysis"}\NormalTok{, }\AttributeTok{cache =} \ConstantTok{TRUE}\NormalTok{)}
\end{Highlighting}
\end{Shaded}

The above R script is saved as \texttt{oser.R}. The central piece of the
executable compendium is the \texttt{Makefile}.

\begin{Shaded}
\begin{Highlighting}[]
\DataTypeTok{output\_file}\CharTok{=}\StringTok{reproduced.html}
\DataTypeTok{r\_cmd }\CharTok{=}\StringTok{ "rmarkdown::render(\textquotesingle{}materials/README.Rmd\textquotesingle{}, output\_file = \textquotesingle{}}\CharTok{$\{}\DataTypeTok{output\_file}\CharTok{\}}\StringTok{\textquotesingle{})"}
\DataTypeTok{handle}\CharTok{=}\StringTok{oser}
\DataTypeTok{local\_file}\CharTok{=$\{}\DataTypeTok{handle}\CharTok{\}}\StringTok{\_README.html}

\DecValTok{all:}\DataTypeTok{ resolve build render}
\NormalTok{    echo }\StringTok{"finished"}
\DecValTok{resolve:}
\NormalTok{    Rscript }\CharTok{$\{}\DataTypeTok{handle}\CharTok{\}}\NormalTok{.R}
\DecValTok{build:}\DataTypeTok{ }\CharTok{$\{}\DataTypeTok{handle}\CharTok{\}}\DataTypeTok{docker}
\NormalTok{    docker build {-}t }\CharTok{$\{}\DataTypeTok{handle}\CharTok{\}}\NormalTok{img }\CharTok{$\{}\DataTypeTok{handle}\CharTok{\}}\NormalTok{docker}
\DecValTok{render:}
\NormalTok{    docker run {-}d {-}{-}rm {-}{-}name }\StringTok{"}\CharTok{$\{}\DataTypeTok{handle}\CharTok{\}}\StringTok{container"}\NormalTok{ {-}ti }\CharTok{$\{}\DataTypeTok{handle}\CharTok{\}}\NormalTok{img}
\NormalTok{    docker exec }\CharTok{$\{}\DataTypeTok{handle}\CharTok{\}}\NormalTok{container Rscript {-}e }\CharTok{$\{}\DataTypeTok{r\_cmd}\CharTok{\}}
\NormalTok{    docker cp }\CharTok{$\{}\DataTypeTok{handle}\CharTok{\}}\NormalTok{container:/materials/}\CharTok{$\{}\DataTypeTok{output\_file}\CharTok{\}} \CharTok{$\{}\DataTypeTok{local\_file}\CharTok{\}}
\NormalTok{    docker stop }\CharTok{$\{}\DataTypeTok{handle}\CharTok{\}}\NormalTok{container}
\DecValTok{export:}
\NormalTok{    docker save }\CharTok{$\{}\DataTypeTok{handle}\CharTok{\}}\NormalTok{img | gzip \textgreater{} }\CharTok{$\{}\DataTypeTok{handle}\CharTok{\}}\NormalTok{img.tar.gz}
\DecValTok{rebuild:}\DataTypeTok{ }\CharTok{$\{}\DataTypeTok{handle}\CharTok{\}}\DataTypeTok{img.tar.gz}
\NormalTok{    docker load \textless{} }\CharTok{$\{}\DataTypeTok{handle}\CharTok{\}}\NormalTok{img.tar.gz}
\end{Highlighting}
\end{Shaded}

With this \texttt{Makefile}, one can create the Dockerfile with
\texttt{make\ resolve}, build the Docker image with
\texttt{make\ build}, render the RMarkdown file inside the container
with \texttt{make\ render}, export the built Docker image with
\texttt{make\ export}, and rebuild the exported Docker image with
\texttt{make\ rebuild}.

The structure of the entire executable compendium looks like this:

\begin{Shaded}
\begin{Highlighting}[]
\ExtensionTok{Makefile}
\ExtensionTok{oser.R}
\ExtensionTok{meta{-}analysis/}
\ExtensionTok{README.md}
\ExtensionTok{oserdocker/}
\ExtensionTok{oserimg.tar.gz}
\end{Highlighting}
\end{Shaded}

In this executable compendium, only the first four elements are
essential. The directory \texttt{oserdocker} (116 MB) contains cached R
packages, a Dockerfile, and a verbatim copy of the directory
\texttt{meta-analysis/} to be transferred into the Docker image. That
can be regenerated by running \texttt{make\ resolve}. However, having
this directory preserved insures against the situations that some R
packages used in the project were no longer available or any of the
information providers used by \texttt{rang} for resolving the dependency
relationships were not available. (Or in the rare circumstance of
\texttt{rang} is no longer available.)

\texttt{oserimg.tar.gz} (667 MB) is a backup copy of the Docker image.
This can be regenerated by running \texttt{make\ export}. Preserving
this file insures against all the situations mentioned above, but also
the situations of Docker Hub and the software repositories used by the
dockerized operating system being not available. When
\texttt{oserimg.tar.gz} is available, it is possible to run
\texttt{make\ rebuild} and \texttt{make\ render} even without internet
access (provided that Docker and \texttt{make} have been installed
before). Of course, there is still an extremely rare situation where
Docker (the program) itself is no longer available \footnote{We can't
  imagine a world without \texttt{Make}, a tool that has been available
  since 1976.}. However, it is possible to convert the image file for
use on other containerization solutions such as Singularity \footnote{https://docs.sylabs.io/guides/3.0/user-guide/singularity\_and\_docker.html},
if Docker is really not available anymore.

Sharing of research artifacts less than 1G is not as challenging as it
used to be. Zenodo, for example, allows the sharing of 50G of files.
Therefore, sharing of the last two components of the executable
compendium prepared with \texttt{rang} is at least possible on Zenodo
\footnote{The complete version of the executable compendium is available
  from Zenodo: https://doi.org/10.5281/zenodo.7708417}. However, for
data repositories with more restrictions on data size, sharing the
executable compendium without the last two parts could be considered
sufficient. For that, run \texttt{make} will make the default target
\texttt{all} and generate all the things needed for reproducing the
analysis inside a container.

The above \texttt{Makefile} is general enough that one can reuse it by
just modifying how the R scripts (the \texttt{r\_cmd} variable) in the
\texttt{materials} directory are executed. This can be a starting point
of a standard executable compendium format.

\hypertarget{concluding-remarks}{%
\section{Concluding remarks}\label{concluding-remarks}}

This paper presents \texttt{rang}, a solution to (re)construct R
computational environments based on Docker. As the six examples in
Section 3 show, \texttt{rang} can be used archaeologically to rerun old
code, many of them not executable without the analytic and
reconstruction processes facilitated by \texttt{rang}. These
retrospective use cases demonstrate how versatile \texttt{rang} is.
\texttt{rang} is also helpful for prospective usage, as demonstrated in
Section 4 whereby an executable compendium is created.

There are still many features that we did not mention in this paper.
\texttt{rang} is built with interoperability in mind. As of writing,
\texttt{rang} is interoperable with existing R packages such as
\texttt{renv} and R built-in \texttt{sessionInfo()}. Also, the
\texttt{rang} object can be used for network analysis with R packages
such as \texttt{igraph}.

Computational reproducibility is a complex topic and as in all of these
complex topics, there is no silver bullet (Canon and Younge 2019). All
solutions have their trade-offs. The (re)construction process based on
\texttt{rang} takes notably more time than other solutions because all
packages are compiled from source. \texttt{rang} trades computational
efficiency of this often one-off (re)constructing process for
correctness, backward compatibility and independence from any commercial
backups of software repositories such as MRAN. There are also other
limitations. In the Vignette of \texttt{rang}
(https://cran.r-project.org/web/packages/rang/vignettes/faq.html), we
list all of these limitations as well as possible mitigation.

\hypertarget{references}{%
\section*{References}\label{references}}
\addcontentsline{toc}{section}{References}

\hypertarget{refs}{}
\begin{CSLReferences}{1}{0}
\leavevmode\vadjust pre{\hypertarget{ref-abate:2015:MCR}{}}%
Abate, Pietro, Roberto Di Cosmo, Louis Gesbert, Fabrice Le Fessant, Ralf
Treinen, and Stefano Zacchiroli. 2015. {``Mining Component Repositories
for Installability Issues.''} \emph{2015 IEEE/ACM 12th Working
Conference on Mining Software Repositories}, May.
\url{https://doi.org/10.1109/msr.2015.10}.

\leavevmode\vadjust pre{\hypertarget{ref-baker:2020:UGM}{}}%
Baker, Peter. 2020. {``Using {GNU Make} to Manage the Workflow of Data
Analysis Projects.''} \emph{Journal of Statistical Software} 94 (Code
Snippet 1). \url{https://doi.org/10.18637/jss.v094.c01}.

\leavevmode\vadjust pre{\hypertarget{ref-beck:2019:EGD}{}}%
Beck, Nathaniel. 2019. {``Estimating Grouped Data Models with a
Binary-Dependent Variable and Fixed Effects via a Logit Versus a Linear
Probability Model: The Impact of Dropped Units.''} \emph{Political
Analysis} 28 (1): 139--45. \url{https://doi.org/10.1017/pan.2019.20}.

\leavevmode\vadjust pre{\hypertarget{ref-benoit:2018}{}}%
Benoit, Kenneth, Kohei Watanabe, Haiyan Wang, Paul Nulty, Adam Obeng,
Stefan Müller, and Akitaka Matsuo. 2018. {``Quanteda: An {R} Package for
the Quantitative Analysis of Textual Data.''} \emph{Journal of Open
Source Software} 3 (30): 774. \url{https://doi.org/10.21105/joss.00774}.

\leavevmode\vadjust pre{\hypertarget{ref-boettiger:2017:IR}{}}%
Boettiger, Carl, and Dirk Eddelbuettel. 2017. {``{An Introduction to
Rocker: Docker Containers for R}.''} \emph{The R Journal} 9 (2): 527.
\url{https://doi.org/10.32614/rj-2017-065}.

\leavevmode\vadjust pre{\hypertarget{ref-canon:2019:CPR}{}}%
Canon, Richard S., and Andrew Younge. 2019. {``A Case for Portability
and Reproducibility of HPC Containers.''} \emph{2019 IEEE/ACM
International Workshop on Containers and New Orchestration Paradigms for
Isolated Environments in HPC (CANOPIE-HPC)}, November.
\url{https://doi.org/10.1109/canopie-hpc49598.2019.00012}.

\leavevmode\vadjust pre{\hypertarget{ref-cruewell:2023:WB}{}}%
Crüwell, Sophia, Deborah Apthorp, Bradley J. Baker, Lincoln Colling,
Malte Elson, Sandra J. Geiger, Sebastian Lobentanzer, et al. 2023.
{``What's in a Badge? A Computational Reproducibility Investigation of
the Open Data Badge Policy in One Issue of Psychological Science.''}
\emph{Psychological Science}, February, 095679762211408.
\url{https://doi.org/10.1177/09567976221140828}.

\leavevmode\vadjust pre{\hypertarget{ref-dolstra:2010:N}{}}%
Dolstra, Eelco, Andres Löh, and Nicolas Pierron. 2010. {``{NixOS}: A
Purely Functional Linux Distribution.''} \emph{Journal of Functional
Programming} 20 (5--6): 577--615.
\url{https://doi.org/10.1017/s0956796810000195}.

\leavevmode\vadjust pre{\hypertarget{ref-hilgard:2019:NEG}{}}%
Hilgard, Joseph, Christopher R. Engelhardt, Jeffrey N. Rouder, Ines L.
Segert, and Bruce D. Bartholow. 2019. {``Null Effects of Game Violence,
Game Difficulty, and 2D:4D Digit Ratio on Aggressive Behavior.''}
\emph{Psychological Science} 30 (4): 606--16.
\url{https://doi.org/10.1177/0956797619829688}.

\leavevmode\vadjust pre{\hypertarget{ref-jurka:2012}{}}%
Jurka, P., Timothy. 2012. {``Maxent: An r Package for Low-Memory
Multinomial Logistic Regression with Support for Semi-Automated Text
Classification.''} \emph{The R Journal} 4 (1): 56.
\url{https://doi.org/10.32614/rj-2012-007}.

\leavevmode\vadjust pre{\hypertarget{ref-kim:2018:E}{}}%
Kim, Yang-Min, Jean-Baptiste Poline, and Guillaume Dumas. 2018.
{``Experimenting with Reproducibility: A Case Study of Robustness in
Bioinformatics.''} \emph{GigaScience} 7 (7).
\url{https://doi.org/10.1093/gigascience/giy077}.

\leavevmode\vadjust pre{\hypertarget{ref-loercher:2017:D}{}}%
Lörcher, Ines, and Monika Taddicken. 2017. {``Discussing Climate Change
Online. Topics and Perceptions in Online Climate Change Communication in
Different Online Public Arenas.''} \emph{Journal of Science
Communication} 16 (02): A03. \url{https://doi.org/10.22323/2.16020203}.

\leavevmode\vadjust pre{\hypertarget{ref-merow:2023:B}{}}%
Merow, Cory, Brad Boyle, Brian J. Enquist, Xiao Feng, Jamie M. Kass,
Brian S. Maitner, Brian McGill, et al. 2023. {``Better Incentives Are
Needed to Reward Academic Software Development.''} \emph{Nature Ecology
\& Evolution}, February.
\url{https://doi.org/10.1038/s41559-023-02008-w}.

\leavevmode\vadjust pre{\hypertarget{ref-nuest:2019}{}}%
Nüst, Daniel, and Matthias Hinz. 2019. {``Containerit: Generating
Dockerfiles for Reproducible Research with {R}.''} \emph{Journal of Open
Source Software} 4 (40): 1603.
\url{https://doi.org/10.21105/joss.01603}.

\leavevmode\vadjust pre{\hypertarget{ref-checkpointrpkg}{}}%
Ooi, Hong, Andrie de Vries, and Microsoft. 2022. \emph{{checkpoint:
Install Packages from Snapshots on the Checkpoint Server for
Reproducibility}}. \url{https://CRAN.R-project.org/package=checkpoint}.

\leavevmode\vadjust pre{\hypertarget{ref-oser:2022:HPE}{}}%
Oser, Jennifer, Amit Grinson, Shelley Boulianne, and Eran Halperin.
2022. {``How Political Efficacy Relates to Online and Offline Political
Participation: A Multilevel Meta-Analysis.''} \emph{Political
Communication} 39 (5): 607--33.
\url{https://doi.org/10.1080/10584609.2022.2086329}.

\leavevmode\vadjust pre{\hypertarget{ref-peikert:2021:RDA}{}}%
Peikert, Aaron, and Andreas M. Brandmaier. 2021. {``A Reproducible Data
Analysis Workflow.''} \emph{Quantitative and Computational Methods in
Behavioral Sciences} 1 (May). \url{https://doi.org/10.5964/qcmb.3763}.

\leavevmode\vadjust pre{\hypertarget{ref-peng:2003:MDP}{}}%
Peng, Roger D. 2003. {``Multi-Dimensional Point Process Models in r.''}
\emph{Journal of Statistical Software} 8 (16).
\url{https://doi.org/10.18637/jss.v008.i16}.

\leavevmode\vadjust pre{\hypertarget{ref-phanstiel:2014:S}{}}%
Phanstiel, D. H., A. P. Boyle, C. L. Araya, and M. P. Snyder. 2014.
{``{Sushi.R}: Flexible, Quantitative and Integrative Genomic
Visualizations for Publication-Quality Multi-Panel Figures.''}
\emph{Bioinformatics} 30 (19): 2808--10.
\url{https://doi.org/10.1093/bioinformatics/btu379}.

\leavevmode\vadjust pre{\hypertarget{ref-groundhogrpkg}{}}%
Simonsohn, Uri, and Hugo Gruson. 2023. \emph{{groundhog: Version-Control
for CRAN, GitHub, and GitLab Packages}}.
\url{https://CRAN.R-project.org/package=groundhog}.

\leavevmode\vadjust pre{\hypertarget{ref-RN-2003-001}{}}%
Tierney, Luke. 2003. {``Name Space Management for {R}.''} \emph{R News}
3: 2--6.

\leavevmode\vadjust pre{\hypertarget{ref-trisovic:2022}{}}%
Trisovic, Ana, Matthew K. Lau, Thomas Pasquier, and Mercè Crosas. 2022.
{``A Large-Scale Study on Research Code Quality and Execution.''}
\emph{Scientific Data} 9 (1).
\url{https://doi.org/10.1038/s41597-022-01143-6}.

\leavevmode\vadjust pre{\hypertarget{ref-renvrpkg}{}}%
Ushey, Kevin. 2022. \emph{{renv: Project Environments}}.
\url{https://CRAN.R-project.org/package=renv}.

\leavevmode\vadjust pre{\hypertarget{ref-valstar:2020:UDS}{}}%
Valstar, Sander, William G. Griswold, and Leo Porter. 2020. {``Using
DevContainers to Standardize Student Development Environments: An
Experience Report.''} \emph{Proceedings of the 2020 ACM Conference on
Innovation and Technology in Computer Science Education}, June.
\url{https://doi.org/10.1145/3341525.3387424}.

\end{CSLReferences}

\end{document}